\documentclass[utf8]{frontiersFPHY} % for Physics and Applied Mathematics and Statistics articles

\setcitestyle{square} % for Physics and Applied Mathematics and Statistics articles
\usepackage{url,hyperref,lineno,microtype,subcaption}
\usepackage[onehalfspacing]{setspace}

\graphicspath{{STA-PSP/}}
\chardef\us=`\_

\def\keyFont{\fontsize{8}{11}\helveticabold }
\def\firstAuthorLast{Hu {et~al.}} %use et al only if is more than 1 author
\def\Authors{Qiang Hu\,$^{1,*}$, Wen He\,$^{1}$, and Yu Chen\,$^{1}$}
\def\Address{$^{1}$Department of Space Science and CSPAR, The University of Alabama in Huntsville, Huntsville, AL, USA}
\def\corrAuthor{Corresponding Author}

\def\corrEmail{qiang.hu@uah.edu}

\begin{document}
\onecolumn
\firstpage{1}

\title[Two-Spacecraft Optimization]{A Magnetic Flux Rope Configuration Derived by Optimization of Two-Spacecraft In-situ Measurements} 

\author[\firstAuthorLast ]{\Authors} %This field will be automatically populated
\address{} %This field will be automatically populated
\correspondance{} %This field will be automatically populated

\extraAuth{}% If there are more than 1 corresponding author, comment this line and uncomment the next one.
%\extraAuth{corresponding Author2 \\ Laboratory X2, Institute X2, Department X2, Organization X2, Street X2, City X2 , State XX2 (only USA, Canada and Australia), Zip Code2, X2 Country X2, email2@uni2.edu}

\maketitle

\begin{abstract}
%
%%%%% Leave the Abstract empty if your article does not require one, please see the Summary Table for full details.
\section{}
Increasingly one interplanetary coronal mass ejection (ICME) structure can propagate across more than one spacecraft in the solar wind. This usually happens when two or more spacecraft are nearly radially aligned with a relatively small longitudinal separation angle from one another. This provides multi-point measurements of the same structure and enables better characterization and validation of modeling results of the structures embedded in these ICMEs. We report such an event during October 13-14, 2019 when the Solar TErrestrial RElations Observatory Ahead (STA) spacecraft  and the Parker Solar Probe (PSP) crossed one ICME structure at two different locations with nominal separations in both heliocentric distances and the longitudinal angles. We first perform an optimal fitting to the STA in-situ measurements, based on an analytic quasi-three dimensional (3D) model, yielding a minimum reduced $\chi^2=0.468$. Then we further apply the optimization approach by combining the magnetic field measurements from both spacecraft along their separate paths across the ICME structure. We find that the output based on the optimization (with the minimum reduced $\chi^2=3.15$) of the combined two-spacecraft dataset yields a  more consistent result, given the much improved agreement of the model output with PSP data. The result demonstrates a magnetic flux rope configuration with clear 3D spatial variations. 
\tiny
 \keyFont{ \section{Keywords:} Magnetic Clouds, Magnetic Flux Ropes, Coronal Mass Ejections, Force-Free Field, Parker Solar Probe} 
 %All article types: you may provide up to 8 keywords; at least 5 are mandatory.
\end{abstract}

\section{Introduction}\label{sec:intro}
Interplanetary coronal mass ejections (ICMEs) are the interplanetary counterparts of CMEs. They propagate into the interplanetary space after CME eruptions and may be detected in-situ by one or more spacecraft en route to larger heliocentric distances away from the Sun. Such in-situ measurements, in the form of time series, present a number of distinctive signatures in both magnetic field and plasma parameters during the passage of an ICME structure \citep{2006SSRv..123...31Z}. They often include, but are not limited to, the elevated magnetic field magnitude and increased solar wind speed, relative to the ambient solar wind immediately preceding the ICME complex. Sometimes the ICME intervals exhibit a specific set of signatures including the elevated magnetic field magnitude, relative smooth rotation in field components, and depressed proton temperature (thus the proton $\beta$, the ratio between the proton thermal pressure and the corresponding magnetic pressure). This type of ICMEs has been traditionally categorized as magnetic clouds (MCs; \citep{1982GeoRL...9.1317B,JA093iA07p07217,1995ISAA....3.....B}).

The internal magnetic structure embedded within an ICME is closely related to its source, the CME eruption on the Sun. It has been increasingly realized that a magnetic flux rope configuration forms the core structure of a CME eruption \citep{Vourlidas2014,2018Natur.554..211A,2021NatAs.tmp..128J}.  In addition to concurrent and modern but indirect observations of flux ropes on the Sun via remote-sensing instrumentations, the in-situ more direct observations of ICME flux ropes have been made by many heliospheric spacecraft missions. These missions include the most recent Parker Solar Probe (PSP; \citep{Fox2016}) and  Solar Orbiter (SO; \citep{SOmission2020}). Both have observed a number of ICME events during their early times at different heliocentric distances, due to their unique orbits around the Sun \cite[see, e.g.,][]{Mostl_2020}\footnote{\url{https://helioforecast.space/icmecat}}. 

These in-situ measurements,  albeit only at a single point or along a single line for one spacecraft, offer perhaps the most quantitative characterization of the ICME structures, thus have enabled a long-lasting effort on modeling the underlying magnetic field configuration based on the in-situ measurements of magnetic field and plasma properties. The earliest and the most representative one is a model fitting approach to a one-dimensional (1D) analytic solution, so-called the Lundquist solution \citep{lund}, based on a linear-force free field (LFFF) assumption. It has been widely applied to a number of spacecraft, ranging from the Interplanetary Monitoring Platform (IMP) \citep{JA086iA08p06673} to the Wind spacecraft \citep[e.g.,][]{Lepping1997}.  The other representative model is the Grad-Shafranov (GS) reconstruction method by solving the two-dimensional (2D) GS equation to obtain a 2D cross section of the magnetic field structure \citep{2001GeoRLHu,2002JGRAHu,Hu2017GSreview}. Both approaches have yielded magnetic flux rope configurations for ICME/MC events examined, but with certain degrees of symmetry, i.e., 1D for the Lundquist solution (only dependence on the radial distance from a central cylindrical axis), and 2D for the GS reconstruction result (arbitrary cross-section geometry with no variation along the cylindrical axis).

The latest development in the ICME flux rope modeling based on in-situ spacecraft measurements takes one step further in showcasing a 3D geometry of the magnetic field configuration.  Based on an LFFF formulation described by \citet{freidberg}, dubbed the Freidberg solution (FS), that includes but is more general than the 1D Lundquist solution by allowing for additional variations in all three spatial dimensions. Hu et al. \citep{2021GeoRL..4890630H} presented the first application of the FS model fitting to in-situ MC measurements. An optimal least-squares fitting based on the standard $\chi^2$ statistics \citep{2002nrca.book.....P,2021Husolphys} was carried out to minimize the $\chi^2$ value between the analytic and measured magnetic field components along a single-spacecraft path subject to measurement uncertainty estimates. The results showed the minimum reduce $\chi^2$ values around 1.0, yielding a set of the corresponding optimal parameters, which characterizes a more general magnetic flux rope configuration with winding flux bundles, displaying topological features likely  corresponding to both ``writhe" and ``twist" of magnetic field lines. The 3D spatial variations are intrinsic to the FS model fitting results.

In addition, important validations to the FS model fitting results were provided through multi-spacecraft studies of CME/ICME flux ropes by using both multi-point in-situ measurements through one ICME structure \citep{2021Husolphys,Hufrontiers2021}, and multi-spacecraft measurements from both in-situ spacecraft crossing an ICME and the corresponding remote-sensing observations of the CME source region \citep{hu2022validation,he2022quantitative}. For example, in \citet{2021Husolphys}, an MC event observed in May 2007 by both STEREO Behind (STB) and the Advanced Composition Explorer (ACE) spacecraft was examined by fitting the FS model to the STB in-situ data. The ACE spacecraft crossed the same solution domain to the west  near ecliptic with respect to STB. The analytic  (``predicted") magnetic field components from the FS model compared well with the corresponding observed ones, along the ACE spacecraft path, yielding an overall correlation coefficient 0.89 between the two sets of data. 

In this study, we continue to perform this type of analysis taking advantage of a rare occasion of one ICME event encountered by both STA and PSP spacecraft with appropriate separations in both heliocentric distances $r_h$, and longitudinal angles. We first carry out a similar analysis as before for such a two-spacecraft encounter by using the STA data only for the FS model fitting. Then we further extend the analysis by incorporating the combined dataset from the two spacecraft into the optimal fitting approach. We demonstrate the merit of the latter approach in terms of overall improved agreement between the FS model result and the magnetic field measurements for both spacecraft along their separate paths. The paper is organized as follows. Section~\ref{sec:twosc} provides a  brief overview of the event and presents the time-series data. Section~\ref{sec:FSfit} describes briefly the analytic FS model fitting, and presents the fitting results for both implementations with STA-only and combined STA-PSP dataset. We summarize and discuss the significance of this type of analysis in Section~\ref{sec:diss}.

\section{A Two-Spacecraft Encounter of an ICME}\label{sec:twosc}
\begin{figure}[h!]
\begin{center}
\includegraphics[width=12cm]{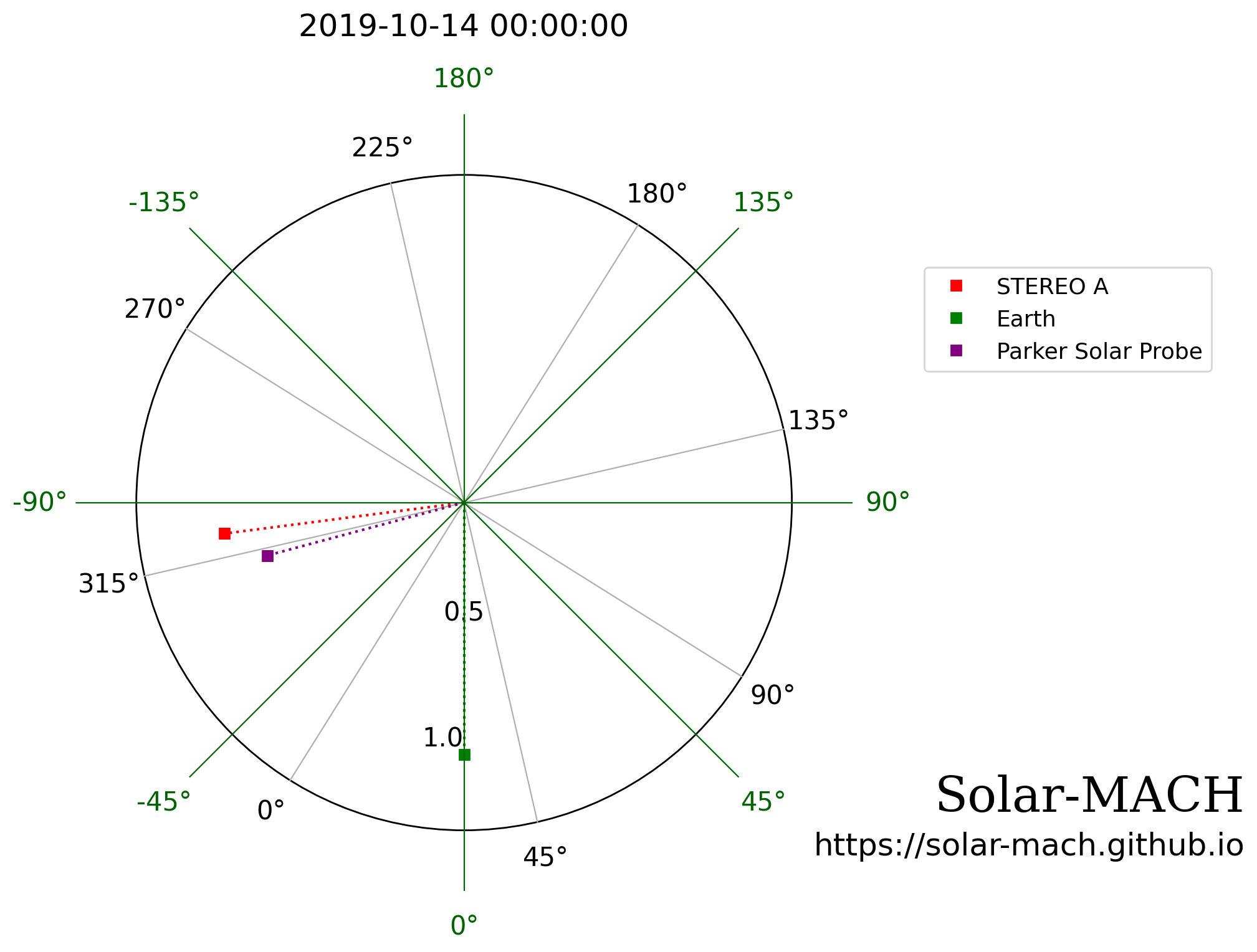}% This is a *.pdf file
\end{center}
\caption{The locations of STA, PSP, and Earth on the equatorial plane on October 14, 2019, as indicated by the legend (generated by the Solar MAgnetic Connection Haus (Solar-MACH) tool; see  {\url{https://da.gd/juUKDZ}}). The black (green) labels are the Carrington (Stonyhurst) longitudes. }\label{fig:1}
\end{figure}

\begin{figure}[h!]
\begin{center}
\includegraphics[width=12cm]{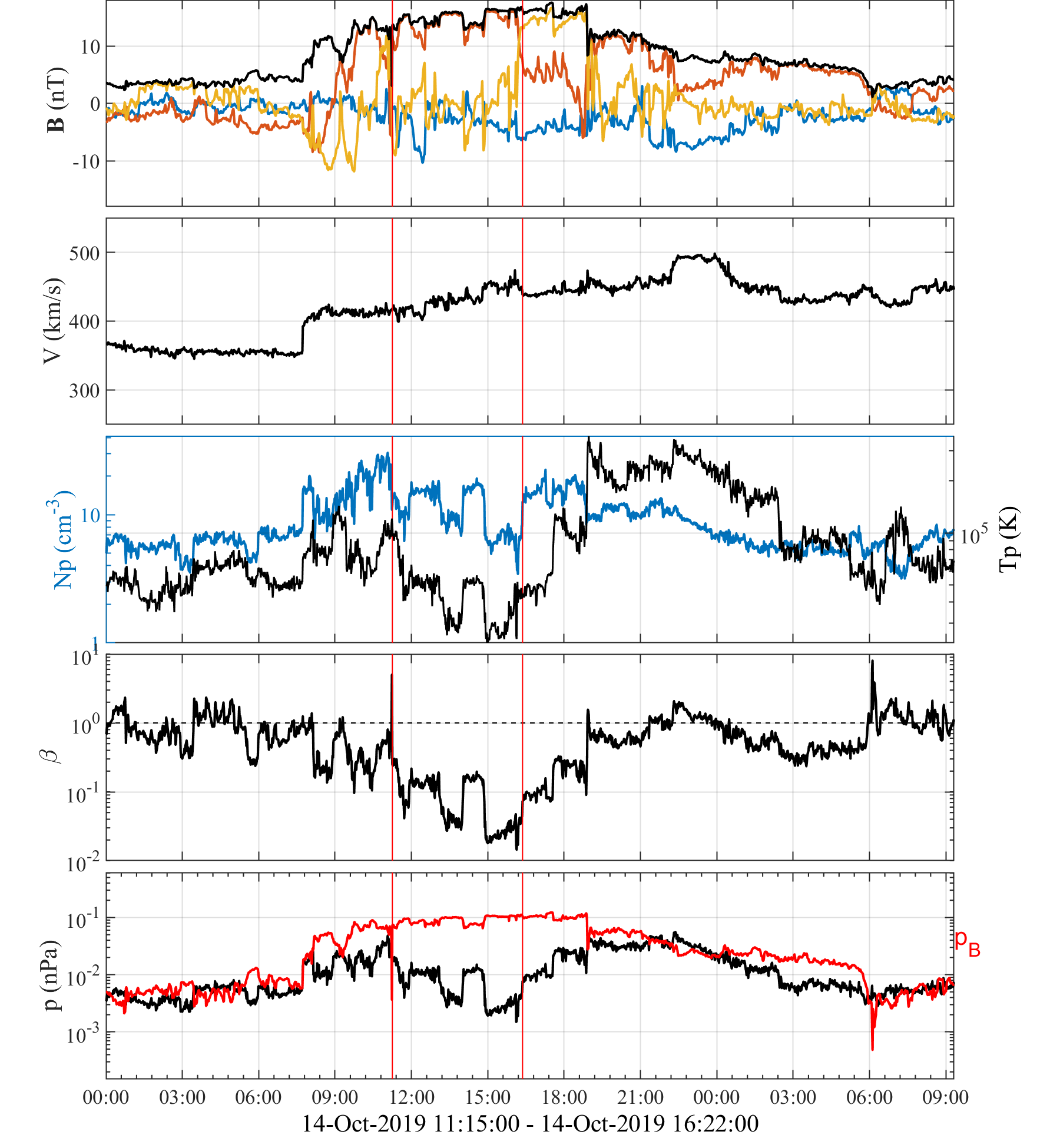}% This is a *.pdf file
\end{center}
\caption{Time-series measurements from the STA spacecraft starting on October 14, 2019. From the top to bottom panels are: the magnetic field components in the Radial (blue), Tangential (red), and Normal (gold) coordinates and magnitude (black),  the solar wind speed, the proton number density (left axis) and temperature (right axis), the proton $\beta$, and the proton pressure and total magnetic pressure $p_B$. The vertical lines mark the interval chosen for analysis, as given beneath the bottom panel in UT. }\label{fig:STAdata}
\end{figure}

An ICME event occurred during October 14-15, 2019, which was observed in-situ by both STA and PSP spacecraft in the heliosphere. Their in-situ measurements were presented in detail by \citet{Winslow_2021}. Figure~\ref{fig:1} shows the locations of the two spacecraft and Earth on the equatorial plane with respect to the Sun in the center. The STA and PSP located at the heliocentric distances $r_h\approx 0.96$ AU and 0.81 AU, respectively. Longitudinally, PSP was separated from STA by about 8$^\circ$ to the west and they also had a relative latitudinal separation of about 1$^\circ$. The time-series measurements \citep{2008SSRv..136..117L,2008SSRv..136..437G} by STA are shown in Figure~\ref{fig:STAdata}, starting on October 14, 2019. The ICME complex extended for nearly a day, driving a shock wave \citep{Winslow_2021}. The signatures in solar wind speed, density, and temperature profiles also indicate the existence of a high-speed stream following the ICME, as discussed in detail by \citet{Winslow_2021}, including the possible solar sources. Here we focus on the analysis of the interval with enhanced magnetic field magnitude and depressed proton temperature in the middle of October 14, 2019.

\begin{figure}[h!]
\begin{center}
\includegraphics[width=15cm]{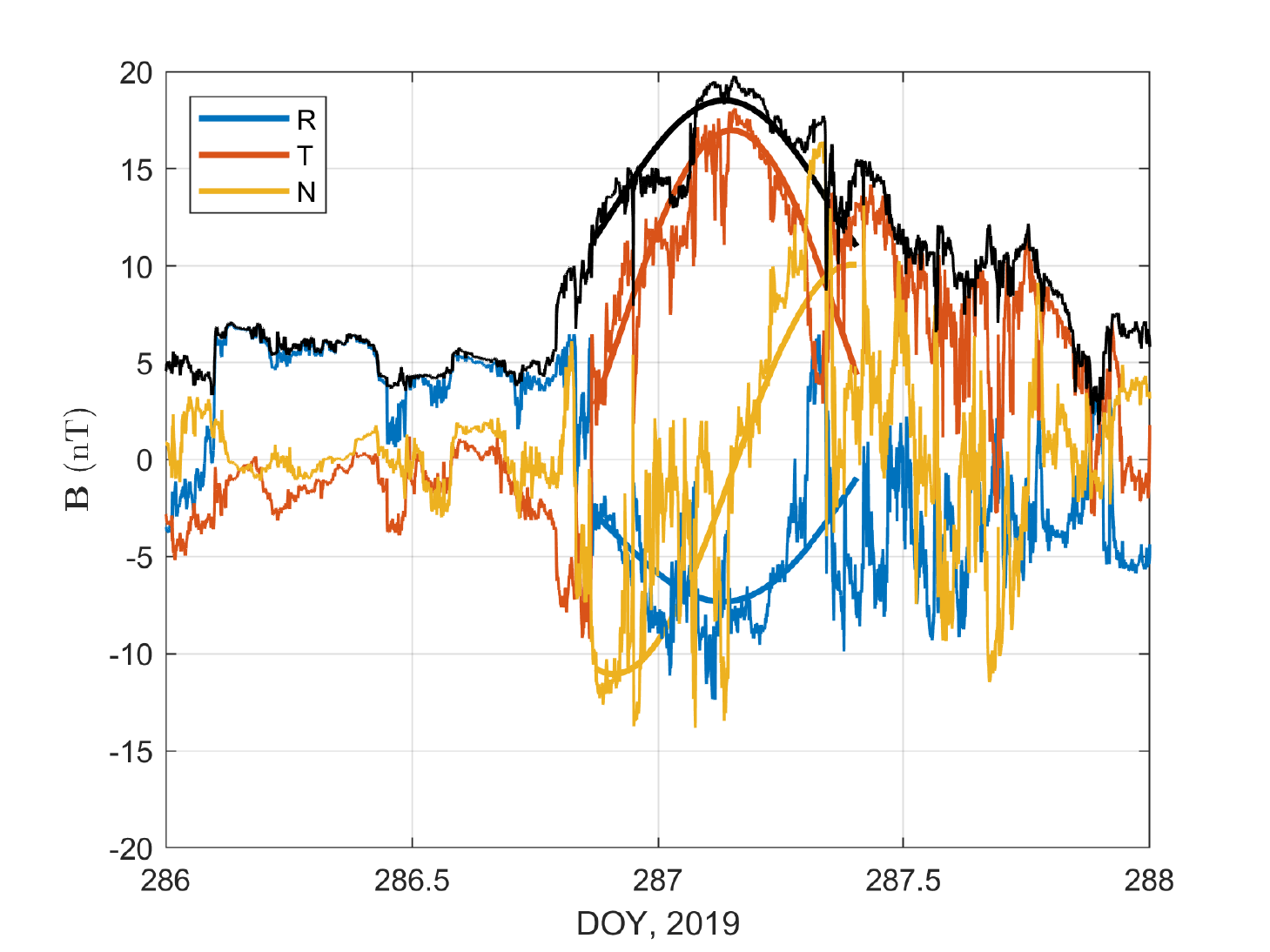}% This is a *.pdf file
\end{center}
\caption{The corresponding magnetic field measurements including the magnitude (black) from PSP in the RTN coordinates during day of year, DOY 286 (October 13), and DOY 288, 2019. The smooth curves are the FS model output from the two-spacecraft optimization presented in Section~\ref{sec:FSfit}.}\label{fig:PSP}
\end{figure}

The vertical lines mark the interval chosen for the FS model fitting to be presented in the next section. During this interval, the field magnitude is relatively high, but rotations of magnetic  field components are not pronounced, especially when compared with the corresponding magnetic field components observed by PSP about half a day earlier (see Figure~\ref{fig:PSP}). The proton $\beta$ is low with an average value 0.092 for the marked interval. The speed shows an increasing profile. The corresponding de Hoffmann-Teller (HT) analysis yields a frame velocity $\mathbf{V}_{HT}=(433.89, 2.78, -4.20)$ km/s in the local Radial, Tangential and Normal (RTN) coordinates (see \citet{2021Husolphys} for the description of and the justification for the HT analysis). This is the velocity with which the magnetic structure propagates across the spacecraft. In other words, the FS model fitting will be performed in this frame of reference with the spacecraft crossing the solution domain with a constant velocity $-\mathbf{V}_{HT}$. The corresponding metric, so-called Wal\'en slope \citep{2021Husolphys}, representing the relative importance of the inertia force over the Lorentz force in the reference frame is 0.027 for this event.  Therefore, strictly speaking, although the interval identified may not correspond to a typical MC interval, the force-free conditions for a static equilibrium  are considered to be satisfied. 

Figure~\ref{fig:PSP} shows the corresponding PSP magnetic field measurements form the FIELDS \citep{2016SSRv..204...49B} instrument (plasma measurements not available), starting on October 13, 2019 (day of year, DOY 286). A significantly different magnetic field profile is seen, compared with Figure~\ref{fig:STAdata}. In the PSP centered RTN coordinates, the N component of the magnetic field has a more pronounced rotation from negative to positive values, and the T component has a fairly symmetric profile with a central peak. The overplotted smooth curves are the corresponding FS model fit, to be presented in Section~\ref{sec:FSfit}.

\section{Modeling and Optimization of the ICME Magnetic Structure}\label{sec:FSfit}
The FS modeling is based on a least-squares minimization of the deviation between the analytic FS model output and the in-situ measured magnetic field components along a spacecraft path via a standard $\chi^2$ optimization approach \citep{2002nrca.book.....P}, as given below, 
\begin{equation}
\chi^2=\frac{1}{\tt{dof}}\sum_{\nu=X,Y,Z}\sum_{i=1}^N\frac{(b_{\nu i} - B_{\nu
i})^2}{\sigma_{ i}^2}. \label{eq:chi2}
\end{equation}
Here the spacecraft measurements are denoted $\mathbf{b}$, and the corresponding results from the analytic FS model are denoted $\mathbf{B}$. All components are given in arrays of length $N$. The uncertainty associated with each data point, $\sigma_i$, is assessed by taking the root-mean-square (RMS) value of each segment of the underlying higher-resolution magnetic field data \citep{2021Husolphys}. For example, in this analysis, the magnetic field data $\mathbf{b}$ are averaged to 20-minute sampling intervals from the corresponding 1-minute resolution data. So each segment is 20 minutes long, containing 21 samples of 1-minute resolution data. The degree of freedom ($\tt{dof}$) is defined as $3N-n-1$, where the number of free parameters is denoted $n$. Thus the reduced $\chi^2$ value is obtained and presented throughout this study. In addition, a metric, representing the quality of ``goodness-of-fit", is calculated as $Q=1 - \tt{chi2cdf}(\chi^2, \tt{dof})$, where the function $\tt{chi2cdf}$ is the cumulative distribution
function of $\chi^2$ \citep{2002nrca.book.....P}. It indicates the probability of the derived minimum $\chi^2$ value being truly the minimum.

The analytic FS model is described by the following set of equations for the three magnetic field components in a cylindrical coordinate system \citep{freidberg}, $(r,\theta,z)$, 
\begin{eqnarray}
\frac{B_z(\mathbf{r})}{B_{z0}} & = & J_0(\mu r)+CJ_1(l r)\cos(\theta+kz), \label{eq:B}\\
\frac{B_\theta(\mathbf{r})}{B_{z0}} & = & J_1(\mu r)-\frac{C}{l}\left[\mu J'_1(l r)+\frac{k}{l r}J_1(l r)\right]\cos(\theta+kz), \\
\frac{B_r(\mathbf{r})}{B_{z0}} & = & -\frac{C}{l}\left[k J'_1(l r)+\frac{\mu}{l r}J_1(l r)\right]\sin(\theta+kz). \label{eq:B4}
\end{eqnarray}
Here the Bessel functions of the first kind of orders 0 and 1 are denoted $J_0$ and $J_1$, respectively. The main constant free parameters are $C$, $k$, and $\mu$, which indicate the amplitude of the non-axisymmetric variation, the wavenumber in the $z$ dimension, and the constant force-free parameter, respectively. Note that for $C\equiv 0$, the axisymmetric Lundquist solution with only $r$ dependence results. Therefore the optimal fitting by the FS model includes and is more general than the Lundquist solution fitting. A few other parameters include $l=\sqrt{\mu^2-k^2}$, and $B_{z0}$ which is pre-determined from the magnetic field measurements as the maximum absolute value among all components for the analysis interval. There is also a to-be-determined length parameter \citep{2021Husolphys}, $a$, with which the parameters $k$ and $\mu$ become dimensionless by multiplying. 

\begin{figure}[h!]
\begin{center}
\includegraphics[width=15cm]{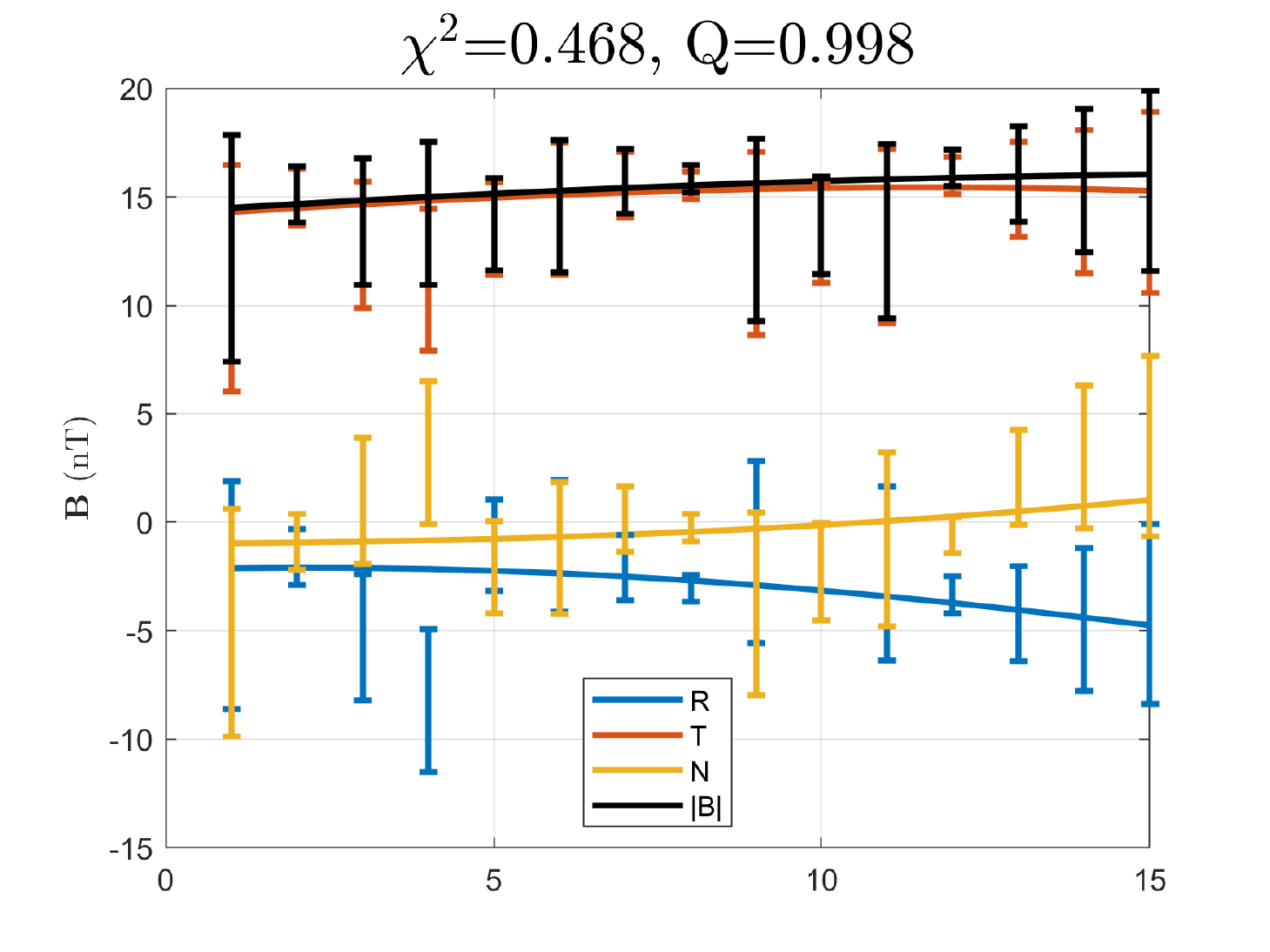}% This is a *.pdf file
\end{center}
\caption{The optimal FS model fitting result (smooth curves) to the STA data (errorbars) only for the interval  marked in  Figure~\ref{fig:STAdata}. The corresponding minimum reduced $\chi^2$ and $Q$ values are denoted on top. }\label{fig:Brtn}
\end{figure}

\begin{figure}[h!]
\begin{center}
\includegraphics[width=15cm]{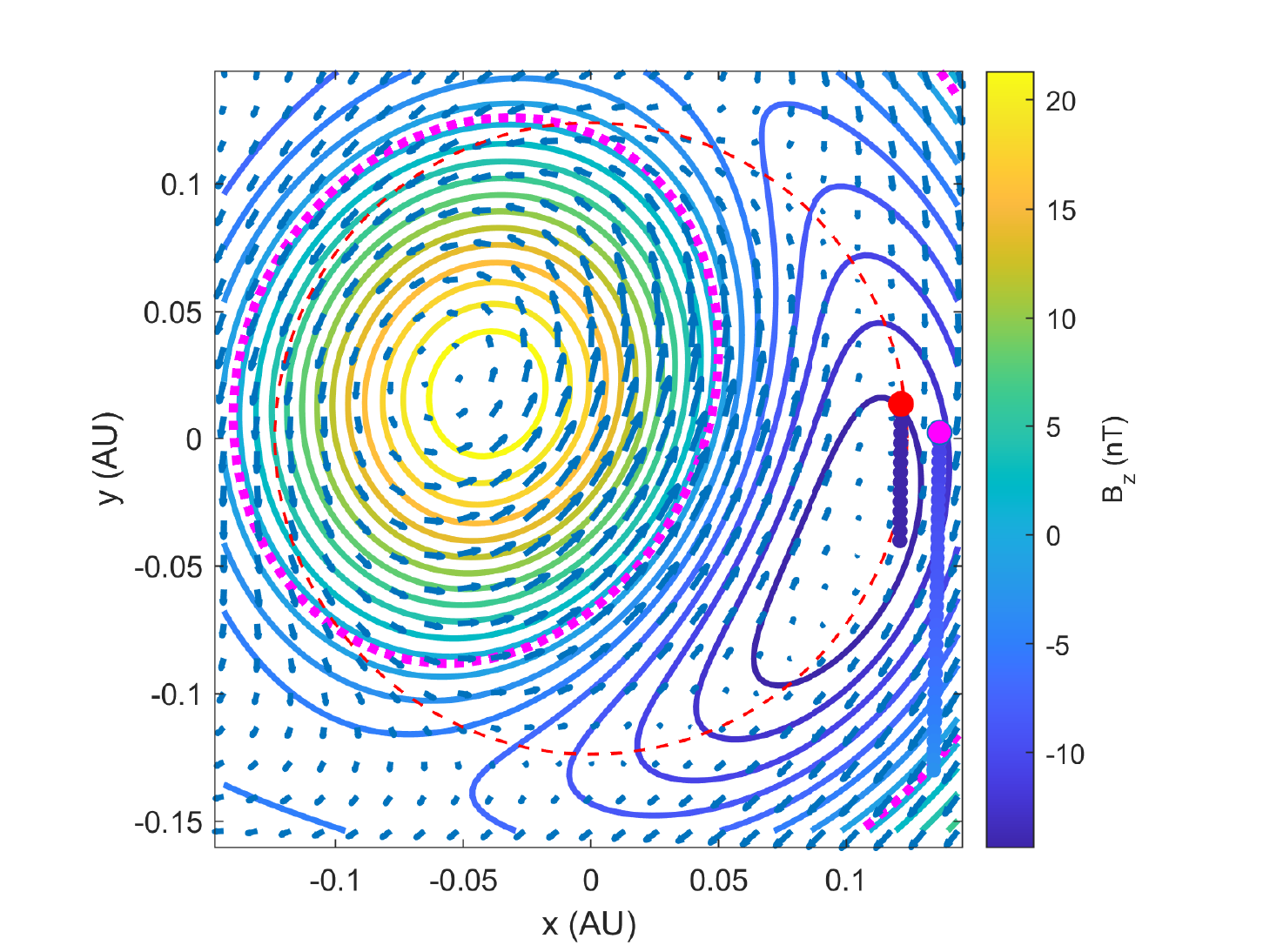}
\end{center}
\caption{One cross section at $z=0$ plane for the optimal FS model corresponding to Figure~\ref{fig:Brtn}. The colored contours are the distribution of $B_z$ with scales given by the colorbar. The magenta dotted curve marks the boundary where $B_z=0$. The blue arrows show the distribution of the transverse magnetic field on this plane. The red thin dashed curve denotes a circle of radius $a=0.12$ AU. The two straight short lines of dots, color-coded by the corresponding $B_z$ values, mark the paths of STA and PSP, with the starting points colored in red and magenta, respectively. }\label{fig:Bz0}
\end{figure}

\begin{figure}[h!]
\begin{center}
\includegraphics[width=15cm]{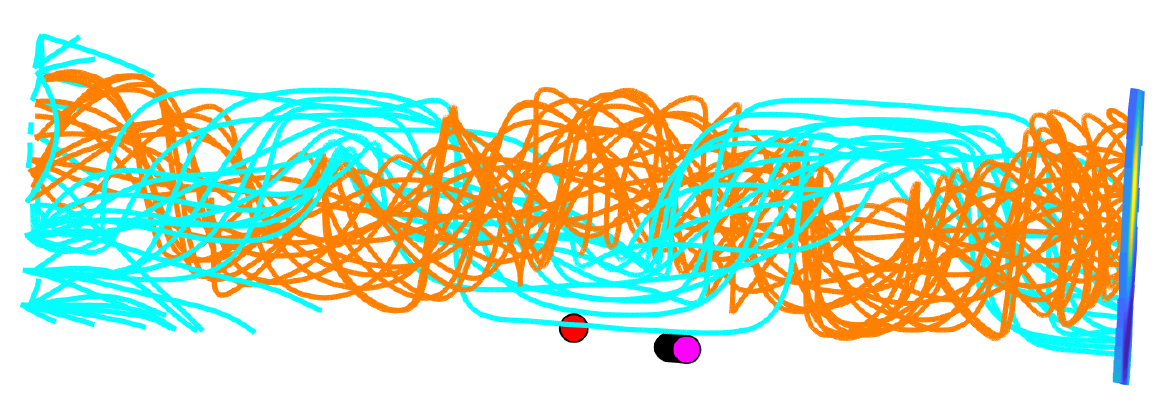}% This is a *.pdf file
\end{center}
\caption{A rendering of the magnetic field lines of the optimal FS model result based on the fitting to the STA-only data interval in a view from the STA's perspective.  The field lines in orange are rooted in the positive $B_z$ polarity region while the cyan lines are rooted on the negative polarity region on the bottom cross section plane which is equivalent to the one given in Figure~\ref{fig:Bz0}. The view is directly along the STA's path (the -R direction) as denoted by the red dots toward the Sun (the N direction is straight up), and the PSP path is marked by the magenta dots to the west.}\label{fig:BfldlineSTA}
\end{figure}

A minimization approach \citep{2021Husolphys} based on equations~(\ref{eq:chi2})-(\ref{eq:B4}) is applied to the STA data only for the interval marked in Figure~\ref{fig:STAdata}. The optimal fitting result is given in Figure~\ref{fig:Brtn} with the corresponding minimum reduced $\chi^2$ and $Q$ values denoted on top. The minimum reduced $\chi^2$ value is less than 1 and the corresponding $Q$ value is nearly 1, indicating a good fitting result. The optimal fitting parameters for the FS model are given in Table~\ref{tbl:case1} with associated uncertainty estimates based on 90\% confidence limits \citep{2002nrca.book.....P,Hufrontiers2021}. A cross section at $z=0$ is shown in Figure~\ref{fig:Bz0}. The solution exhibits two flux bundles of opposite polarities (nonzero and opposite $B_z$ components) next to each other. In other words, the field lines corresponding to the two polarity regions are directed in opposite directions. It is seen that both STA and PSP spacecraft crossed the bundle of negative $B_z$ with the positive $z$ axis oriented in a direction that is nearly aligned with the east-west direction (see Table~\ref{tbl:case1}). The configuration is better visualized in Figure~\ref{fig:BfldlineSTA} where selected field lines rooted in both positive and negative polarity regions are drawn and viewed from the STA's perspective toward the Sun. The flux bundle in cyan corresponds to field lines rooted in the negative polarity region shown in Figure~\ref{fig:Bz0} and is crossed by both STA and PSP spacecraft. They are  pointing toward the west and wrapping around the other flux bundle in orange color. Both are winding along the $z$ axis with the orientation angles given in Table~\ref{tbl:case1}, but pointing in opposite directions. The axial magnetic flux is estimated  for the negative polarity region over the cross section and its magnitude with uncertainty is given in Table~\ref{tbl:case1}.

\begin{figure}[h!]
\begin{center}
\includegraphics[width=\textwidth]{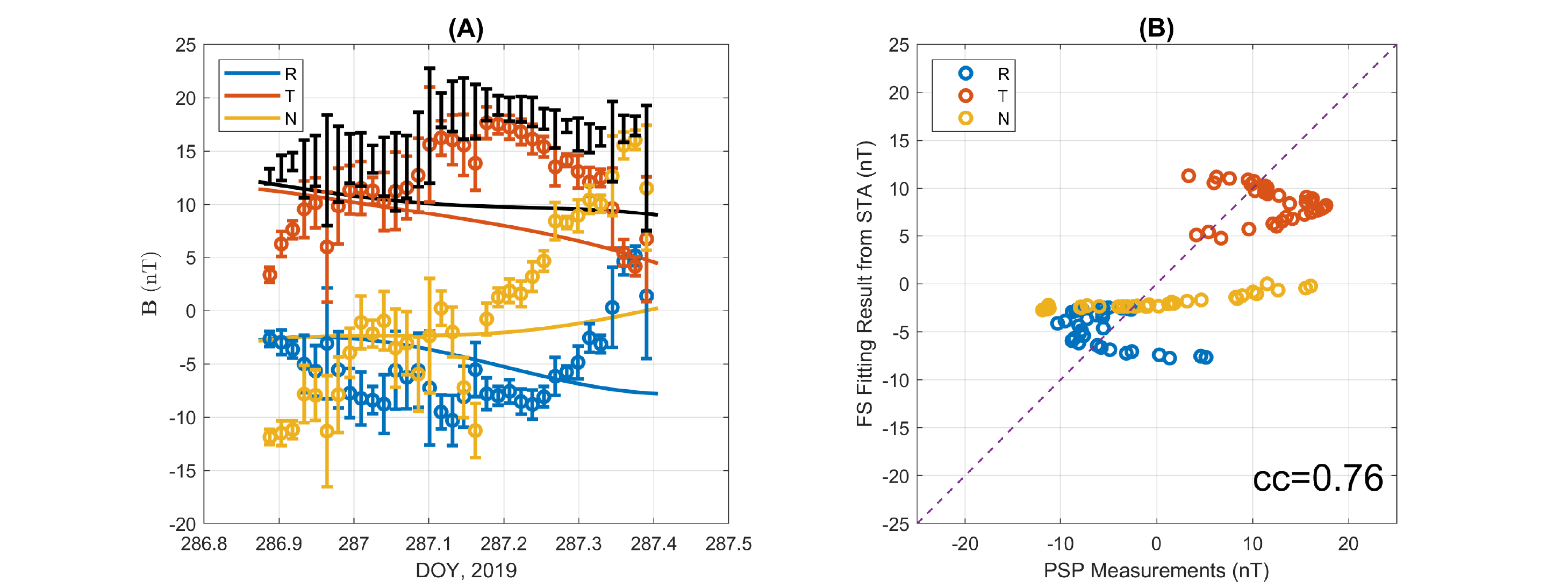}% This is a *.pdf file
\end{center}
\caption{(A) The comparison between the FS model output (smooth curves) based on the optimal fitting to STA data only and the actual measurements (errorbars), along the PSP spacecraft path, as illustrated in Figure~\ref{fig:BfldlineSTA}. (B) The corresponding component-wise correlation plot between the two sets of data, yielding an overall correlation coefficient $cc=0.76$. The diagonal dashed line indicates the one-to-one line. The corresponding correlation coefficients for  each magnetic field component in the RTN coordinates are  -0.46, -0.14, and 0.91, respectively. }\label{fig:corrSTA}
\end{figure}

It is a useful practice to compare the magnetic field components, along the PSP spacecraft path, yielded by the  FS model fit to the STA data only with the actual PSP measurements. Such a comparison is given in Figure~\ref{fig:corrSTA}, after taking into account a nominal time shift due to the separation in $r_h$ and a constant propagating speed $|\mathbf{V}_{HT}|$. The result indicates a poor agreement between the model ``predicted" and the actually measured magnetic field components. The matching in either the rotations of the field components or their magnitudes is lacking.
 
This leads to the next step of the analysis in order to improve the consistency of the FS model result with both spacecraft datasets. It seems feasible, given that the fitting to STA data alone has yielded such a small optimal $\chi^2$ value. Hence we combine the two-spacecraft datasets into one vector field $\mathbf{b}$ in equation~(\ref{eq:chi2}), and carry out the minimization approach, using the two separate and distinct spacecraft paths across the solution domain. This process involves combined  data points which are not continuous and are not along the same straight line. Other than that, the minimization algorithm is the same as applied for the STA-only data.

\begin{figure}[h!]
\begin{center}
\hspace{8.5cm}(B)\vspace{-.7cm}
\includegraphics[width=.48\textwidth]{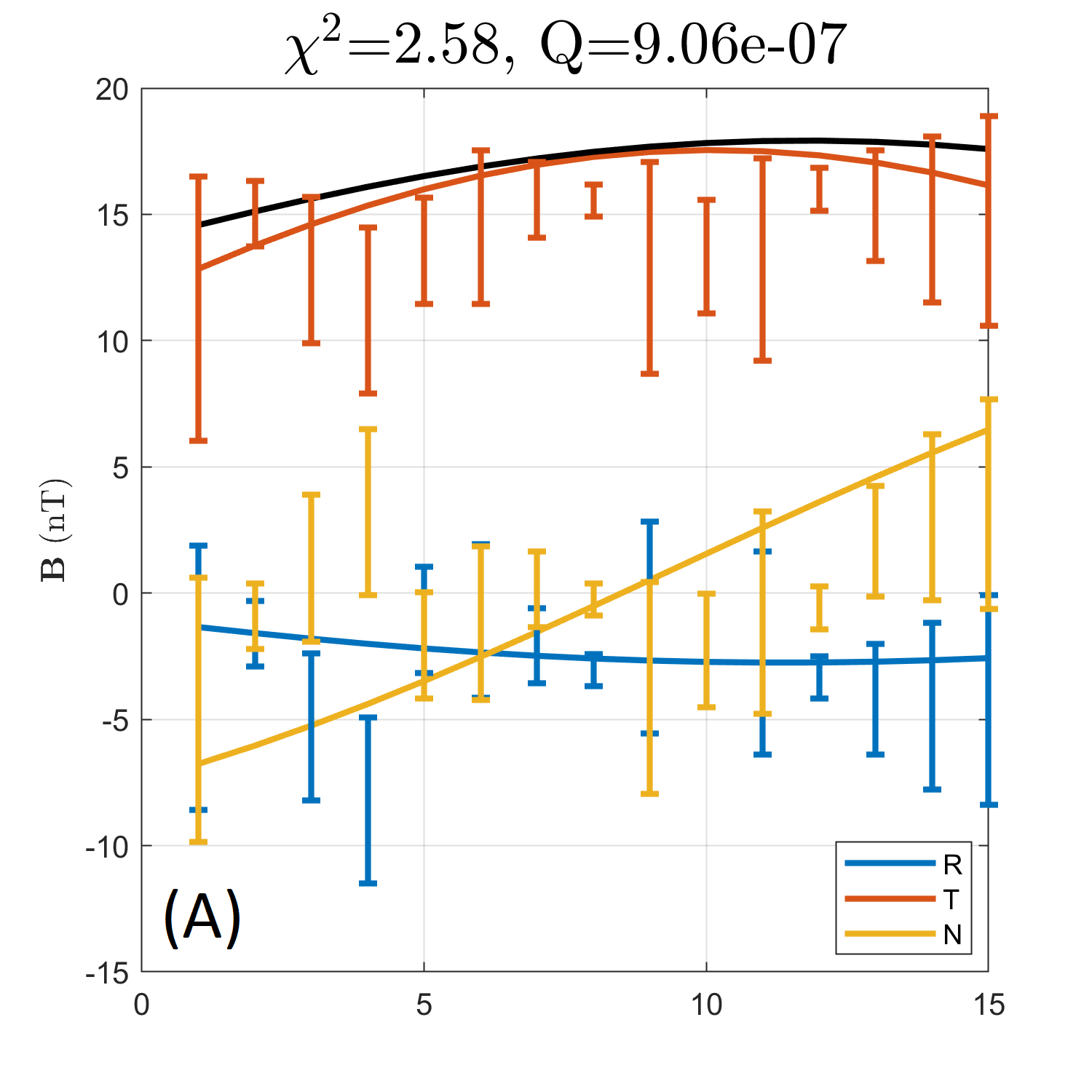}% This is a *.pdf file
\includegraphics[width=.52\textwidth]{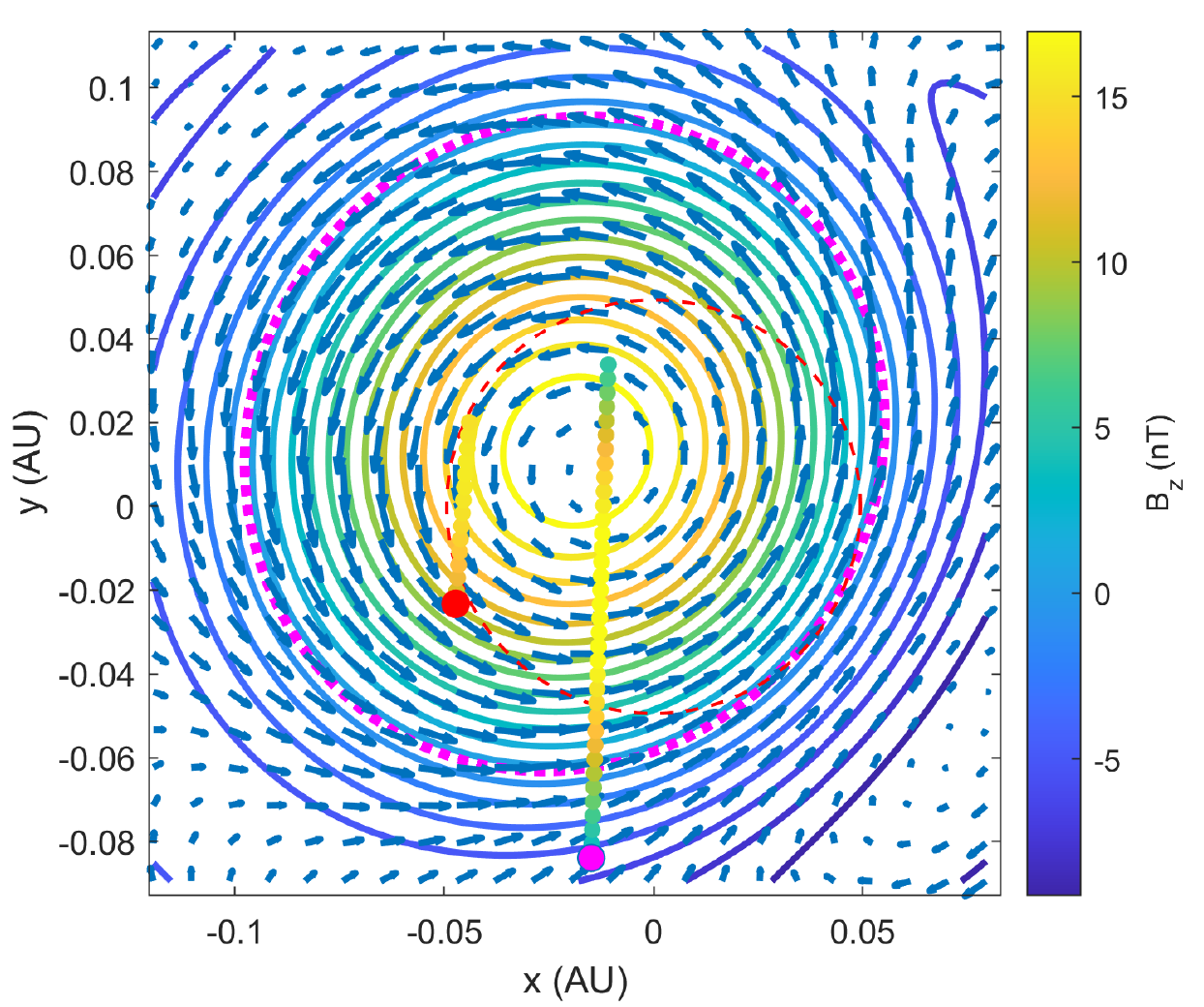}
%\hspace{-4.5cm}(A)\vspace{-1.7cm}
\end{center}
\caption{(A) The two-spacecraft (STA-PSP) optimization result of the magnetic field components and magnitude (black curve) for the STA data interval. Format is the same as Figure~\ref{fig:Brtn}. (B) One cross section for the optimal FS model result based on the two-spacecraft (STA-PSP) optimization (with $a=0.049$ AU). Format is the same as Figure~\ref{fig:Bz0}.  }\label{fig:chi2_2sc}
\end{figure}

Figure~\ref{fig:chi2_2sc} shows the optimization result through the process by combining the two-spacecraft datasets. Figure~\ref{fig:chi2_2sc}A shows the model output compared with the data points along the STA path only. The corresponding minimum reduced $\chi^2$ value increases to 2.58. The three fitted curves to the field components show little variation, except for the $B_N$ component.  Figure~\ref{fig:chi2_2sc}B shows a cross section in the same format as Figure~\ref{fig:Bz0}. The main difference from the previous STA-only solution is that there exists only one dominant polarity in the current solution. The PSP spacecraft path crosses near the ``center", in this view, along which the axial field $B_z$ reaches peak values near the middle of the path. Toward both ends of the PSP path, the axial field $B_z$ decreases significantly, as indicated by the colorbar. The axial magnetic flux is estimated for the positive polarity region over the cross section and  is given in Table~\ref{tbl:case1}. Note that in this solution, the $z$ axis orientation is nearly reversed with respect to the previous STA-only solution, as given in Table~\ref{tbl:case1}. The minimum reduced $\chi^2$ value for the combined STA-PSP two-spacecraft dataset (a total of 49  data points) is 3.15. 

\begin{figure}[h!]
\begin{center}
\includegraphics[width=\textwidth]{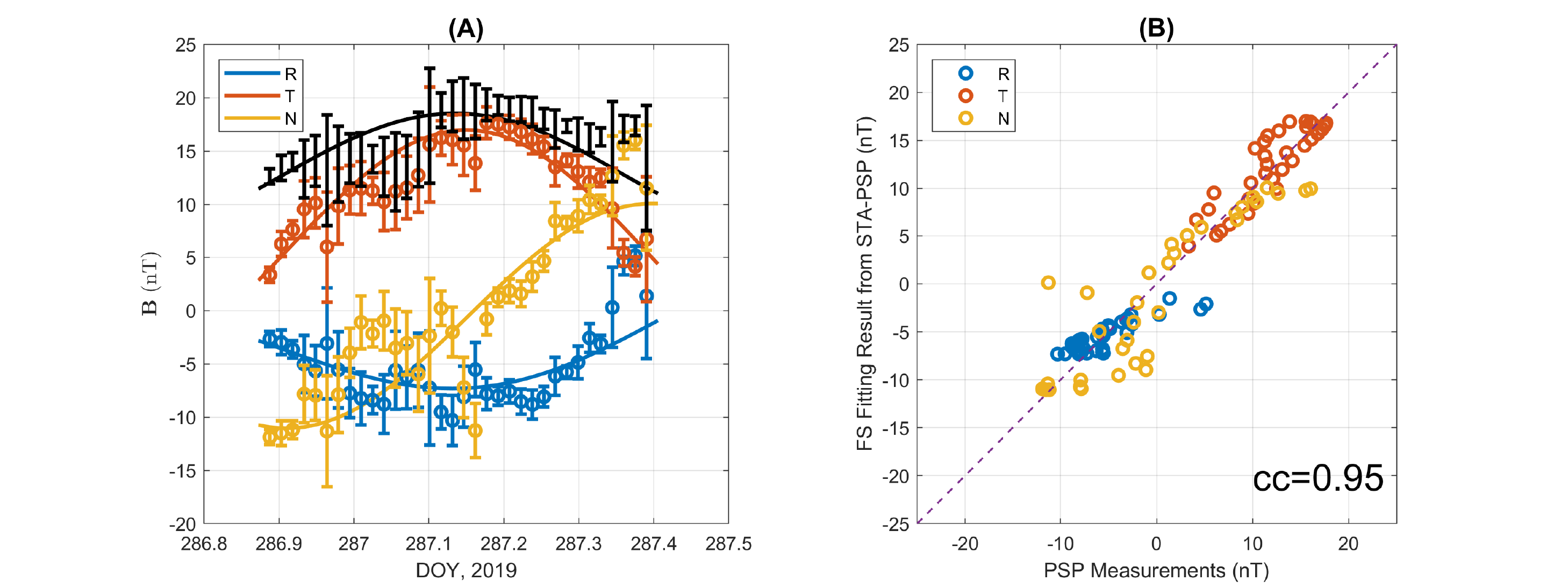}% This is a *.pdf file
\end{center}% STA-PSP
\caption{ The comparison between the FS model output  based on the optimal fitting to the combined STA-PSP dataset  and the actual measurements  along the PSP spacecraft path. Format is the same as Figure~\ref{fig:corrSTA}.  The corresponding correlation coefficients for  each magnetic field component in the RTN coordinates are  0.87, 0.89, and 0.89, respectively, for this analysis result. }\label{fig:corr2sc}
\end{figure}

\begin{figure}[h!]
\begin{center}
\includegraphics[width=15cm]{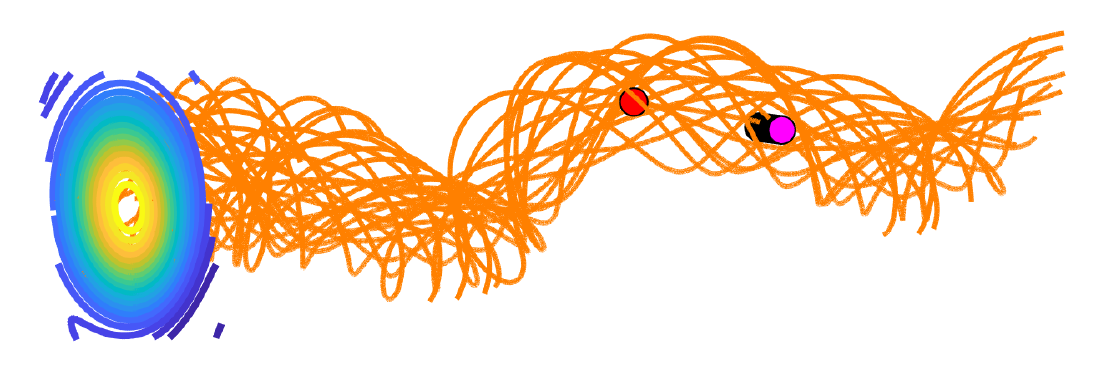}% This is a *.pdf file
\end{center}
\caption{The corresponding field line configuration for the STA-PSP optimal fitting result in a 3D view along the STA path toward the Sun. Format is the same as Figure~\ref{fig:BfldlineSTA}. Note here there exists only one major positive $B_z$ polarity on the bottom plane shown to the left (east with respect to the Sun).  }\label{fig:Bfldline2sc}
\end{figure}

\begin{figure}[h!]
\begin{center}
\includegraphics[width=15cm]{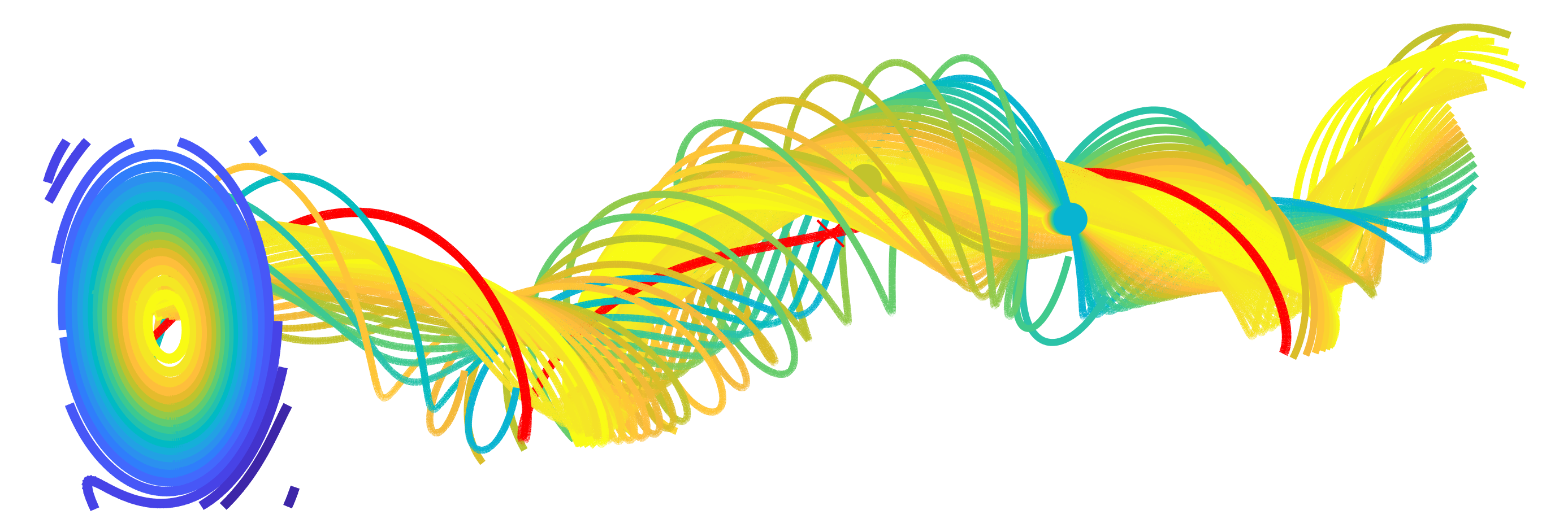}% This is a *.pdf file
\end{center}
\caption{In exactly the same view as Figure~\ref{fig:Bfldline2sc}, selected field lines crossing the spacecraft paths of STA and PSP, and color coded by the corresponding $B_z$ values according to the colorbar of Figure~\ref{fig:chi2_2sc}B.  The thick red line originates from the point with the maximum  $B_z$ on the bottom plane.  }\label{fig:Bfldline2scOrb}
\end{figure}

Figure~\ref{fig:corr2sc} shows the comparison between the FS model output based on the combined STA-PSP dataset  and the actual measurements by PSP along its path as illustrated in Figure~\ref{fig:chi2_2sc}B. The overall correlation coefficient for all three components reaches 0.95, while the correlation coefficients for each individual component are all close to 0.9. Both the peak near the middle of the $B_T$ component and the significant rotation in the $B_N$ component are well recovered by the corresponding FS model output. Figure~\ref{fig:Bfldline2sc} shows, in the same view as Figure~\ref{fig:BfldlineSTA}, the 3D field lines originating from the bottom plane and corresponding to the major positive polarity. One flux bundle is winding in the approximately east-west direction with clear features of writhe or overall winding of the body of the flux bundle in orange color. The two spacecraft are seen to cross the flux bundle at two different locations. The 3D nature of the solution gives rise to the significant difference between the times series returned by the two spacecraft, owing to the spatial variations as revealed by this analysis result. To further illustrate such variations, Figure~\ref{fig:Bfldline2scOrb} shows, in exactly the same view and coordinates, the selected field lines crossing the two spacecraft paths and color-coded by the corresponding $B_z$ values at the intercepting points on each path. They are mostly positive, indicating that all the field lines drawn are pointing to the west. Overall they exhibit a ``twisted-ribbon" type of topology without a central straight field line, a feature that has commonly been found in FS model results \citep{2021GeoRL..4890630H,2021Husolphys,Hufrontiers2021,hu2022validation}. For instance, the thick red line originates from the ``center" with the maximum $B_z$ value on one cross-section plane, but it is not straight due to the variation along the $z$ dimension.  It is also worth noting that the flux bundle in cyan color in Figure~\ref{fig:BfldlineSTA} is pointing in grossly the same direction as the main flux bundle shown here in orange color.  However, the detailed configurations are different as indicated by both the appearances from STA's perspective demonstrated so far, as well as the derived sets of optimal model parameters presented in Table~\ref{tbl:case1}.

\section{Summary and Discussion}\label{sec:diss}
In summary, we have carried out a unique analysis of an ICME magnetic flux rope structure by employing the combined dataset from both STA and PSP spacecraft along their separate paths across the same structure. The results show that the optimization approach based on the usual  $\chi^2$ minimization via the FS model and the two-spacecraft dataset yields a much improved agreement between the model-predicted and the actual measured magnetic field components along the PSP spacecraft path which was away from STA by $\sim8^\circ$ in longitude to the west. The overall correlation coefficient for all three components reaches 0.95, as opposed to the corresponding  value 0.76 from the optimization result based on STA in-situ data alone. This analysis further demonstrates the importance and necessity for employing multi-spacecraft measurements in quantitatively  examining the ICME structures. 

The model result confirms a 3D magnetic flux rope configuration with spiral field lines forming a main flux bundle that exhibits significant winding itself, extending in approximately the east-west direction from STA's perspective. Such a configuration clearly possesses more complex 3D spatial variations, intrinsic to the FS model. It results in the significantly different time series returned by STA and PSP, respectively, because of the different paths across the flux bundle, as seen in Figure~\ref{fig:Bfldline2sc}. In addition, the 3D features are further illustrated in Figure~\ref{fig:Bfldline2scOrb} in that in addition to the lack of symmetry, no straight field lines exist in such a magnetic field configuration and the overall appearance exhibits a ``twisted-ribbon" type of geometry.

It is also worth noting that the combined two-spacecraft optimization is still largely based on or best to start from the single-spacecraft analysis by using the FS model. For this event, the initial analysis by using the STA data alone yields a minimum reduced $\chi^2\approx 0.468$ which lends confidence to the subsequent optimization by employing the two-spacecraft dataset in order to yield  acceptable results  for  both spacecraft and improved consistency. The results from STA-only analysis also show limited consistency with the two-spacecraft optimization result. For instance, in addition to the same chirality (handedness) of the magnetic field topology,  the overall orientation of the flux bundle crossed by both spacecraft also points in the approximate east-west direction. However  the axial magnetic flux content differs. This is due to the significantly different cross-section shape of the negative polarity region given in Figure~\ref{fig:Bz0} which is not well bounded by a closed boundary either, as compared with the positive polarity region given in Figure~\ref{fig:chi2_2sc}B for the STA-PSP fitting result.   In practice, both approaches should be applied, although the two-spacecraft optimization is expected to yield more consistent results for better validation and interpretation, as demonstrated in this event study. Based on the current analysis with the FS model-based optimization and validation by the two-spacecraft measurements, it also provides certain guidance on the range of acceptable minimum reduced $\chi^2$ values which may be extended to as large as 2 to 3 for an optimal fitting to a single-spacecraft dataset. But again it will be essential to validate the results by employing multi-spacecraft measurements whenever available.
% FLUX ?

One may image that a future mission with the goal  of returning multi-point measurements as demonstrated in this study is desirable for modeling ICME and other large-scale structures. It could be composed of two or more probes with one being the primary one carrying a comprehensive set of instruments designated for both magnetic field and plasma measurements. The other or others can serve as ``sidekicks" to the main one, maintaining  appropriate separation distances. Some may only need to carry a set of magnetometers to enable  additional measurements of magnetic field only, which will aid in the modeling of encountered ICME structures by combining multi-point datasets, as we have demonstrated here. One feasible solution is to make use of the existing spacecraft constellations near Earth - ``Spaceship Earth", such as ACE and Wind, as the primary probe, supplemented with the secondary ones as described above to accomplish the goal. Clearly as more data points are obtained, the complexity and generality of the underlying analytic model has to improve, which demands a constantly ongoing effort. 

\section*{Conflict of Interest Statement}
%All financial, commercial or other relationships that might be perceived by the academic community as representing a potential conflict of interest must be disclosed. If no such relationship exists, authors will be asked to confirm the following statement: 

The authors declare that the research was conducted in the absence of any commercial or financial relationships that could be construed as a potential conflict of interest.

\section*{Author Contributions}
QH performed the analysis and prepared the manuscript. YC and WH provided time-series data and   contributed to the final editing and proofreading of the manuscript.

\section*{Funding}
Funding is provided, in part, by NASA grants 80NSSC21K0003, 80NSSC19K0276, 80NSSC18K0622, 80NSSC17K0016, and NSF grants AGS-1650854 and AGS-1954503, to The University of Alabama in Huntsville.

\section*{Acknowledgments}
WH and QH acknowledge NSO/NSF DKIST Ambassador program for support.

\section*{Supplemental Data}

\section*{Data Availability Statement}
The datasets analyzed for this study can be found in the NASA CDAWeb: {\url{https://cdaweb.gsfc.nasa.gov/index.html/}}.
\bibliographystyle{frontiersinHLTH&FPHY} % for Health, Physics and Mathematics articles
\bibliography{ref_master3}

\begin{thebibliography}{27}
\expandafter\ifx\csname natexlab\endcsname\relax\def\natexlab#1{#1}\fi
\expandafter\ifx\csname urlstyle\endcsname\relax
  \expandafter\ifx\csname doi\endcsname\relax
  \def\doi#1{doi:\discretionary{}{}{}#1}\fi \else
  \expandafter\ifx\csname doi\endcsname\relax
  \def\doi{doi:\discretionary{}{}{}\begingroup \urlstyle{rm}\Url}\fi \fi
\expandafter\ifx\csname selectlanguage\endcsname\relax
  \def\selectlanguage#1{}\fi

\bibitem[{{Zurbuchen} and {Richardson}(2006)}]{2006SSRv..123...31Z}
{Zurbuchen} TH, {Richardson} IG.
\newblock {In-Situ Solar Wind and Magnetic Field Signatures of Interplanetary
  Coronal Mass Ejections}.
\newblock {\em \ssr\/} {\bf 123} (2006) 31--43.
\newblock \doi{10.1007/s11214-006-9010-4}.

\bibitem[{{Burlaga} et~al.(1982){Burlaga}, {Klein}, {Sheeley}, {Michels},
  {Howard}, {Koomen} et~al.}]{1982GeoRL...9.1317B}
{Burlaga} LF, {Klein} L, {Sheeley} J N~R, {Michels} DJ, {Howard} RA, {Koomen}
  MJ, et~al.
\newblock {A magnetic cloud and a coronal mass ejection}.
\newblock {\em \grl\/} {\bf 9} (1982) 1317--1320.
\newblock \doi{10.1029/GL009i012p01317}.

\bibitem[{Burlaga(1988)}]{JA093iA07p07217}
Burlaga LF.
\newblock Magnetic clouds and force-free fields with constant alpha.
\newblock {\em Journal of Geophysical Research: Space Physics\/} {\bf 93}
  (1988) 7217--7224.
\newblock \doi{https://doi.org/10.1029/JA093iA07p07217}.

\bibitem[{{Burlaga}(1995)}]{1995ISAA....3.....B}
{Burlaga} LF.
\newblock {Interplanetary magnetohydrodynamics.}
\newblock {\em Interplanetary magnetohydrodynamics, by L.~F.~Burlag.~
  International Series in Astronomy and Astrophysics, Vol.~3, Oxford University
  Press.~1995.~272 pages; ISBN13: 978-0-19-508472-6\/} {\bf 3} (1995).

\bibitem[{{Vourlidas}(2014)}]{Vourlidas2014}
{Vourlidas} A.
\newblock {The flux rope nature of coronal mass ejections}.
\newblock {\em Plasma Physics and Controlled Fusion\/} {\bf 56} (2014) 064001.
\newblock \doi{10.1088/0741-3335/56/6/064001}.

\bibitem[{{Amari} et~al.(2018){Amari}, {Canou}, {Aly}, {Delyon}, and
  {Alauzet}}]{2018Natur.554..211A}
{Amari} T, {Canou} A, {Aly} JJ, {Delyon} F, {Alauzet} F.
\newblock {Magnetic cage and rope as the key for solar eruptions}.
\newblock {\em \nat\/} {\bf 554} (2018) 211--215.
\newblock \doi{10.1038/nature24671}.

\bibitem[{{Jiang} et~al.(2021){Jiang}, {Feng}, {Liu}, {Yan}, {Hu}, {Moore}
  et~al.}]{2021NatAs.tmp..128J}
{Jiang} C, {Feng} X, {Liu} R, {Yan} X, {Hu} Q, {Moore} RL, et~al.
\newblock {A fundamental mechanism of solar eruption initiation}.
\newblock {\em Nature Astronomy\/}  (2021).
\newblock \doi{10.1038/s41550-021-01414-z}.

\bibitem[{Fox et~al.(2016)Fox, Velli, Bale, Decker, Driesman, Howard
  et~al.}]{Fox2016}
Fox N, Velli M, Bale S, Decker R, Driesman A, Howard R, et~al.
\newblock The solar probe plus mission: humanity’s first visit to our star.
\newblock {\em Space Science Reviews\/} {\bf 204} (2016) 7--48.

\bibitem[{{M\"uller, D.} et~al.(2020){M\"uller, D.}, {St. Cyr, O. C.},
  {Zouganelis, I.}, {Gilbert, H. R.}, {Marsden, R.}, {Nieves-Chinchilla, T.}
  et~al.}]{SOmission2020}
{M\"uller, D}, {St Cyr, O C}, {Zouganelis, I}, {Gilbert, H R}, {Marsden, R},
  {Nieves-Chinchilla, T}, et~al.
\newblock The solar orbiter mission - science overview.
\newblock {\em A\&A\/} {\bf 642} (2020) A1.
\newblock \doi{10.1051/0004-6361/202038467}.

\bibitem[{Möstl et~al.(2020)Möstl, Weiss, Bailey, Reiss, Amerstorfer,
  Hinterreiter et~al.}]{Mostl_2020}
Möstl C, Weiss AJ, Bailey RL, Reiss MA, Amerstorfer T, Hinterreiter J, et~al.
\newblock Prediction of the in situ coronal mass ejection rate for solar cycle
  25: Implications for parker solar probe in situ observations.
\newblock {\em The Astrophysical Journal\/} {\bf 903} (2020) 92.
\newblock \doi{10.3847/1538-4357/abb9a1}.

\bibitem[{{Lundquist}(1950)}]{lund}
{Lundquist} S.
\newblock {On force-free solution}.
\newblock {\em Ark. Fys.\/} {\bf 2} (1950) 361.

\bibitem[{Burlaga et~al.(1981)Burlaga, Sittler, Mariani, and
  Schwenn}]{JA086iA08p06673}
Burlaga L, Sittler E, Mariani F, Schwenn R.
\newblock Magnetic loop behind an interplanetary shock: Voyager, helios, and
  imp 8 observations.
\newblock {\em Journal of Geophysical Research: Space Physics\/} {\bf 86}
  (1981) 6673--6684.
\newblock \doi{https://doi.org/10.1029/JA086iA08p06673}.

\bibitem[{{Lepping} et~al.(1997){Lepping}, {Burlaga}, {Szabo}, {Ogilvie},
  {Mish}, {Vassiliadis} et~al.}]{Lepping1997}
{Lepping} RP, {Burlaga} LF, {Szabo} A, {Ogilvie} KW, {Mish} WH, {Vassiliadis}
  D, et~al.
\newblock {The Wind magnetic cloud and events of October 18-20, 1995:
  Interplanetary properties and as triggers for geomagnetic activity}.
\newblock {\em \jgr\/} {\bf 102} (1997) 14049--14064.
\newblock \doi{10.1029/97JA00272}.

\bibitem[{{Hu} and {Sonnerup}(2001)}]{2001GeoRLHu}
{Hu} Q, {Sonnerup} BU{\"O}.
\newblock {Reconstruction of magnetic flux ropes in the solar wind}.
\newblock {\em \grl\/} {\bf 28} (2001) 467--470.
\newblock \doi{10.1029/2000GL012232}.

\bibitem[{{Hu} and {Sonnerup}(2002)}]{2002JGRAHu}
{Hu} Q, {Sonnerup} BU{\"O}.
\newblock {Reconstruction of magnetic clouds in the solar wind: Orientations
  and configurations}.
\newblock {\em \jgra\/} {\bf 107} (2002) 1142.
\newblock \doi{10.1029/2001JA000293}.

\bibitem[{{Hu}(2017)}]{Hu2017GSreview}
{Hu} Q.
\newblock {The Grad-Shafranov Reconstruction in Twenty Years: 1996 - 2016}.
\newblock {\em Sci.~China Earth Sciences\/} {\bf 60} (2017) 1466--1494.
\newblock \doi{doi: 10.1007/s11430-017-9067-2}.

\bibitem[{Freidberg(2014)}]{freidberg}
Freidberg JP.
\newblock {\em Ideal MHD\/} (Cambridge, UK: Cambridge University Press) (2014),
  546--547.

\bibitem[{{Hu} et~al.(2021){Hu}, {He}, {Qiu}, {Vourlidas}, and
  {Zhu}}]{2021GeoRL..4890630H}
{Hu} Q, {He} W, {Qiu} J, {Vourlidas} A, {Zhu} C.
\newblock {On the Quasi-Three Dimensional Configuration of Magnetic Clouds}
  {\bf 48} (2021) e90630.
\newblock \doi{10.1029/2020GL090630}.

\bibitem[{{Press} et~al.(2007){Press}, {Teukolsky}, {Vetterling}, and
  {Flannery}}]{2002nrca.book.....P}
{Press} WH, {Teukolsky} SA, {Vetterling} WT, {Flannery} BP.
\newblock {\em {Numerical Recipes in C++ : The Art of Scientific Computing}\/}
  (New York: 778, Cambridge Univ. Press) (2007).
\newblock \doi{http://numerical.recipes/}.

\bibitem[{Hu(2021)}]{2021Husolphys}
Hu Q.
\newblock Optimal fitting of the freidberg solution to in situ spacecraft
  measurements of magnetic clouds.
\newblock {\em Sol. Phys.\/}  (2021) https://arxiv.org/abs/2104.09352.
\newblock \doi{10.1007/s11207-021-01843-z}.

\bibitem[{Hu et~al.(2021)Hu, He, Zhao, and Lu}]{Hufrontiers2021}
Hu Q, He W, Zhao L, Lu E.
\newblock Configuration of a magnetic cloud from solar orbiter and wind
  spacecraft in-situ measurements.
\newblock {\em Frontiers in Physics\/} {\bf 9} (2021) 407.
\newblock \doi{10.3389/fphy.2021.706056}.

\bibitem[{{Hu} et~al.(2022){Hu}, {Zhu}, {He}, {Qiu}, {Jian}, and
  {Prasad}}]{hu2022validation}
{Hu} Q, {Zhu} C, {He} W, {Qiu} J, {Jian} LK, {Prasad} A.
\newblock {Validation and interpretation of three-dimensional configuration of
  a magnetic cloud flux rope}.
\newblock {\em arXiv e-prints\/}  (2022) arXiv:2204.03457.

\bibitem[{{He} et~al.(2022){He}, {Hu}, {Jiang}, {Qiu}, and
  {Prasad}}]{he2022quantitative}
{He} W, {Hu} Q, {Jiang} C, {Qiu} J, {Prasad} A.
\newblock {Quantitative Characterization of Magnetic Flux Rope Properties for
  Two Solar Eruption Events}.
\newblock {\em arXiv e-prints\/}  (2022) arXiv:2201.03149.

\bibitem[{Winslow et~al.(2021)Winslow, Lugaz, Scolini, and
  Galvin}]{Winslow_2021}
Winslow RM, Lugaz N, Scolini C, Galvin AB.
\newblock First simultaneous in situ measurements of a coronal mass ejection by
  parker solar probe and {STEREO}-a.
\newblock {\em The Astrophysical Journal\/} {\bf 916} (2021) 94.
\newblock \doi{10.3847/1538-4357/ac0821}.

\bibitem[{{Luhmann} et~al.(2008){Luhmann}, {Curtis}, {Schroeder}, {McCauley},
  {Lin}, {Larson} et~al.}]{2008SSRv..136..117L}
{Luhmann} JG, {Curtis} DW, {Schroeder} P, {McCauley} J, {Lin} RP, {Larson} DE,
  et~al.
\newblock {STEREO IMPACT Investigation Goals, Measurements, and Data Products
  Overview}.
\newblock {\em \ssr\/} {\bf 136} (2008) 117--184.
\newblock \doi{10.1007/s11214-007-9170-x}.

\bibitem[{{Galvin} et~al.(2008){Galvin}, {Kistler}, {Popecki}, {Farrugia},
  {Simunac}, {Ellis} et~al.}]{2008SSRv..136..437G}
{Galvin} AB, {Kistler} LM, {Popecki} MA, {Farrugia} CJ, {Simunac} KDC, {Ellis}
  L, et~al.
\newblock {The Plasma and Suprathermal Ion Composition (PLASTIC) Investigation
  on the STEREO Observatories}.
\newblock {\em \ssr\/} {\bf 136} (2008) 437--486.
\newblock \doi{10.1007/s11214-007-9296-x}.

\bibitem[{{Bale} et~al.(2016){Bale}, {Goetz}, {Harvey}, {Turin}, {Bonnell},
  {Dudok de Wit} et~al.}]{2016SSRv..204...49B}
{Bale} SD, {Goetz} K, {Harvey} PR, {Turin} P, {Bonnell} JW, {Dudok de Wit} T,
  et~al.
\newblock {The FIELDS Instrument Suite for Solar Probe Plus. Measuring the
  Coronal Plasma and Magnetic Field, Plasma Waves and Turbulence, and Radio
  Signatures of Solar Transients}.
\newblock {\em \ssr\/} {\bf 204} (2016) 49--82.
\newblock \doi{10.1007/s11214-016-0244-5}.

\end{thebibliography}

%%% Make sure to upload the bib file along with the tex file and PDF
%%% Please see the test.bib file for some examples of references

\section*{Figure captions}

%%% Please be aware that for original research articles we only permit a combined number of 15 figures and tables, one figure with multiple subfigures will count as only one figure.
%%% Use this if adding the figures directly in the mansucript, if so, please remember to also upload the files when submitting your article
%%% There is no need for adding the file termination, as long as you indicate where the file is saved. In the examples below the files (logo1.eps and logos.eps) are in the Frontiers LaTeX folder
%%% If using *.tif files convert them to .jpg or .png
%%%  NB logo1.eps is required in the path in order to correctly compile front page header %%%
%Bz0=16
{
\begin{table}[tbh]
\centering
    \caption{Summary of geometrical and physical  parameters for the ICME/MC from the FS model fitting  to the STA-only and combined STA-PSP spacecraft in-situ measurements.}\label{tbl:case1}
    \begin{tabular}{cccccccc}
        \hline
        %\specialrule{1.5pt}{1pt}{1pt} % Thick line.
        Parameters& $\chi^2$  & $C$& $\mu$ & $k$ & $\hat{\mathbf{z}}=(\delta, \phi)^1$ &  $\Phi_z$ (Mx)& Chirality\\
\hline\hline
        FS (STA-only) & 0.468 & -1.68 & 3.18 & 1.42 &  $(87, 272)$   &7.8 - 9.4  & $\mathbf{+}$\\
                      &            & $\pm0.52$   &  $\pm0.24$   & $\pm0.59$  &  $\pm(15, 15)$             &   $\times10^{20}$    &(right-handed) \\ %90\% conf. limits R0=.12
\hline
    FS (STA-PSP) &3.15 & 0.885 &1.58    &-0.683 & (83,127)   & 2.7 - 3.9  & $\mathbf{+}$\\
                         &     &$\pm0.55 $&$\pm0.52$ &$\pm0.22$&$\pm (8,7)$  & $\times10^{20}$ &(right-handed) \\ %R0=0.0494, chi2=3.15
    \hline
    \end{tabular}\\ $^1$\small{The polar angle $\delta$ from the N direction, and the azimuthal angle $\phi$ measured from R towards T axes in the STA centered RTN coordinates, all in degrees.}
\end{table}}
%
%<\beta>
%C_k_mu = -1.6777    1.4156    3.1830

%https://www.esrl.noaa.gov/gmd/grad/neubrew/Calendar.jsp
%https://www.scp.byu.edu/docs/doychart.html

%ccRTN:  -0.4600   -0.1363    0.9102

%ccRTN:      0.8707    0.8871    0.8911

% add a plot showing field lines crossing the s/c paths

%%% If you are submitting a figure with subfigures please combine these into one image file with part labels integrated.
%%% If you don't add the figures in the LaTeX files, please upload them when submitting the article.
%%% Frontiers will add the figures at the end of the provisional pdf automatically
%%% The use of LaTeX coding to draw Diagrams/Figures/Structures should be avoided. They should be external callouts including graphics.

\end{document}